# Security Authentication as Protective Friction: Understanding the Dual Effects of Authentication Measures in e-Government Platform

Jianshan Pan[1*],
Zongyi Chen[2],
Qiyou Lin[3],
Xiaofang Cai[4*]

*Corresponding author
[1] Director, Shenzhen Public Credit Center, Shenzhen, China, panjs@mail.amr.sz.gov.cn
[2] Associate Director, Shenzhen Public Credit Center, Shenzhen, China, chenzy@mail.amr.sz.gov.cn
[3] Department Head, Shenzhen Public Credit Center, Shenzhen, China, *linqy1@mail.amr.sz.gov.cn*
[4] Associate Researcher, National Center of Applied mathematics, Southern University of Science and Technology, Shenzhen, China, caixf@mail.sustech.edu.cn

## ABSTRACT

E-Government platforms, such as government credit platforms, handle highly sensitive personal data, requiring stringent security authentication (e.g., facial recognition, multi-factor verification) that protects users but impedes their experience. Drawing on Tomalin's (2023) e-friction typology and the emerging literature on beneficial design friction, we reconceptualize these authentication measures as protective friction, a theoretically distinct form of non-elective, overt, impeding e-friction that serves a security-enhancing function. We propose a dual-pathway model in which protective friction simultaneously reduces privacy-security risk perception (protection pathway) and increases effort cost (cost pathway), with self-efficacy moderating only the risk pathway. This differential pattern suggests that what matters is not users' ability to overcome friction, but their comprehension of its protective purpose, a construct we term protective mechanism comprehension. We outline a mixed-methods research design combining a survey of platform users and semi-structured interviews with platform staff at a Chinese municipal credit center to test this model. This research-in-progress contributes to IS theory by extending e-friction research from conceptual typology to empirical investigation, identifying when security friction becomes beneficial rather than burdensome.

## INTRODUCTION

A growing body of technology adoption research has examined how perceived risks and perceived value influence users' intention to adopt digital government services (Carter & Belanger, 2005; Lallmahomed et al., 2017; Sharma et al., 2024). However, this literature typically treats security measures as contextual background rather than as an active design element that shapes user experience. Security authentication, such as facial recognition, SMS verification, and identity document upload, is generally assumed to reduce risk perceptions; its potential to simultaneously increase adoption barriers through added complexity receives less systematic attention.

Recent work in information systems and human-computer interaction has begun to challenge the dominant frictionless paradigm. Tomalin (2023), writing in the Journal of Information Technology, develops a typology of e-friction and argues that protective forms of friction are ethically desirable. Cox et al. (2016) introduce the concept of microboundaries, small, intentional obstacles that disrupt mindless interactions and promote reflection. Distler et al. (2020) propose a framework of security-enhancing friction that integrates UX considerations with security objectives. Together, these contributions suggest that the relationship between friction and user outcomes is more nuanced than the traditional assumption that all friction is harmful. Recent work on nudge and sludge by Lee and Kim (2023) further elucidates how intentional friction can guide better decision-making when properly designed.

In the context of government credit platforms operating within China's social credit system, the governance dynamics are particularly complex. Wang et al. (2023) and Liu and Rona-Tas (2025) document how credit systems institutionalize moral standards and trust mechanisms within this distinctive institutional context. These platforms thus carry unique cultural and institutional significance that shapes user perceptions of authentication friction.

This research-in-progress addresses a gap at the intersection of these literatures by developing and proposing an empirical test of a dual-pathway model of protective friction in the context of government credit platform adoption. We hypothesize that authentication-based protective friction simultaneously (1) reduces privacy-security risk perception through a protection pathway and (2) increases effort cost through a cost pathway. Furthermore, we propose that self-efficacy moderates these pathways differentially, moderating the risk pathway but not the cost pathway, suggesting that the critical factor is users' comprehension of friction's protective purpose rather than their operational ability to overcome it.

This study contributes to the literature in three ways. First, it extends the emerging friction literature from conceptual typology to empirical hypothesis testing in an IS adoption context. Second, it proposes a dual-pathway mechanism through which a single design element (authentication) produces opposing effects on adoption. Third, it identifies a potential construct gap, protective mechanism comprehension, that existing constructs (self-efficacy, perceived effectiveness, security awareness) may fail to capture, opening a new line of inquiry for IS researchers studying security-sensitive digital services.

## THEORETICAL BACKGROUND

### Online Friction: From Elimination to Intentional Design

The technology industry has long operated under the assumption that friction, any impediment to seamless user flow, should be minimized or eliminated. However, a growing body of research challenges this frictionless paradigm. Cox et al. (2016), in their influential CHI workshop paper, introduce the concept of design frictions and microboundaries, small, intentional obstacles that disrupt mindless automatic interactions and promote moments of reflection. They argue that such frictions can reduce errors in data-entry tasks, support health-behavior change, and encourage more deliberate decision-making. This foundational work established that friction is not inherently negative but can be a purposeful design tool.

Building on this insight, Mejtoft et al. (2019) empirically tested the effects of intentionally added design friction on user satisfaction in a mobile application context. Their study at the European Conference on Cognitive Ergonomics found that participants preferred the application with added design friction and reported higher satisfaction when they had a clear understanding of the task's goal, an early hint that friction's effect depends on users' comprehension of its purpose. This finding resonates with research on processing fluency in cognitive psychology, which demonstrates that disfluent (harder-to-process) stimuli can prompt deeper, more analytical thinking rather than superficial responses (Alter et al., 2007).

In the specific domain of security, Distler et al. (2020) propose the Framework of Security-Enhancing Friction, presented at the New Security Paradigms Workshop. Their framework integrates user experience (UX) research with usable security, arguing that friction can be systematically designed to help users behave more securely. Crucially, they emphasize that the experiential quality of security friction, not merely its presence or absence, determines whether it enhances or undermines security behavior. Similarly, Peer et al. (2020) demonstrate that security nudges are more effective when personalized to users' decision-making styles, further supporting the view that user characteristics moderate friction's effects. The broader nudge and sludge framework, analyzed meta-analytically by Lee and Kim (2023), demonstrates that intentional obstacles can improve decision quality when users understand their purpose.

Tomalin (2023) develops a hierarchical typology of e-friction, defined as any unnecessary retardation of a process or activity that delays the user accomplishing a desired action. His framework classifies e-friction along several binary dimensions: elective versus non-elective, overt versus covert, intended versus unintended, impeding versus distracting, and protective versus commercial. His central argument is that protective e-friction, friction that shields users from harm, is ethically desirable and should not be indiscriminately eliminated. He cites Microsoft's multi-factor authentication as an example. However, his analysis remains purely conceptual; he offers no empirical evidence for how users actually experience and respond to protective friction, nor does he theorize the mechanisms through which protective friction affects technology adoption.

In the context of government credit platforms, security authentication measures (facial recognition, multi-factor verification, identity document validation) constitute a specific subtype of e-friction that is: (1) non-elective (users must comply); (2) overt (users are aware of the steps); (3) intended (deliberately designed); (4) impeding (blocking access until completed); and (5) protective (safeguarding sensitive credit information). This classification positions authentication as a prime empirical case for testing the theoretical claims emerging from the friction literature.

### The Dual-Pathway Model of Protective Friction

We propose that protective friction operates through two simultaneous and opposing pathways. The protection pathway: by requiring rigorous identity verification, the platform signals that it takes data security seriously, thereby reducing users' privacy-security risk perception. This aligns with signaling theory, where visible security measures serve as signals of trustworthiness (Mavlanova et al., 2012; Shin, 2024). The cost pathway: the same authentication steps increase the time, effort, and cognitive resources required to complete a task, raising the perceived effort cost. This is consistent with the security-enhancing friction framework (Distler et al., 2020) and research on authentication burden in digital services. The net effect on adoption intention depends on the relative strength of these two opposing pathways.

### Self-Efficacy and the Comprehension Question

Self-efficacy, an individual's belief in their ability to perform a specific task (Bandura, 1977), has been widely used as a moderator in technology adoption research. Recent work on digital self-efficacy by Paredes-Aguirre et al. (2024) has further refined measurement approaches. The standard assumption is that high self-efficacy users are better equipped to overcome technology barriers. If this ability mechanism were the primary driver, self-efficacy should more strongly moderate the effort-cost pathway, since effort cost directly reflects operational difficulty. However, an alternative comprehension mechanism is possible: self-efficacy may proxy for users' technical understanding, enabling them to comprehend why protective friction exists and to evaluate its necessity rationally rather than react with anxiety. Under this interpretation, self-efficacy would more strongly moderate the risk-perception pathway. The pattern of moderation across the two pathways provides a diagnostic test between these competing mechanisms.

**Technology Adoption in Government Digital Services**

Our model integrates established constructs from the technology adoption literature. On the value side, we draw on the Technology Acceptance Model (Davis, 1989) and the DeLone and McLean IS Success Model (DeLone & McLean, 2003) to include perceived usefulness, system quality, information quality, and service accessibility as positive predictors of adoption intention. Research by AlAwadhi et al. (2024) on trust-building factors in e-government and Mensah and Mwakapesa (2025) on information quality dimensions further substantiate the importance of these constructs. On the risk side, we draw on perceived risk theory (Featherman & Pavlou, 2003) and recent work on privacy calculus in digital government contexts (Huang & Huang, 2023; Meier & Kramer, 2024) to include privacy-security risk, effort cost, and financial risk. The government credit platform context introduces a distinctive feature: the institutional trust buffer. Because the platform is government-endorsed and provides free services, financial risk perception may be neutralized, a pattern documented in prior e-government adoption research (Carter & Belanger, 2005).

## RESEARCH MODEL AND HYPOTHESES

Based on the theoretical background above, we propose the research model shown in Figure 1. The model integrates the dual-pathway mechanism of protective friction with established value and risk predictors of technology adoption, moderated by self-efficacy.

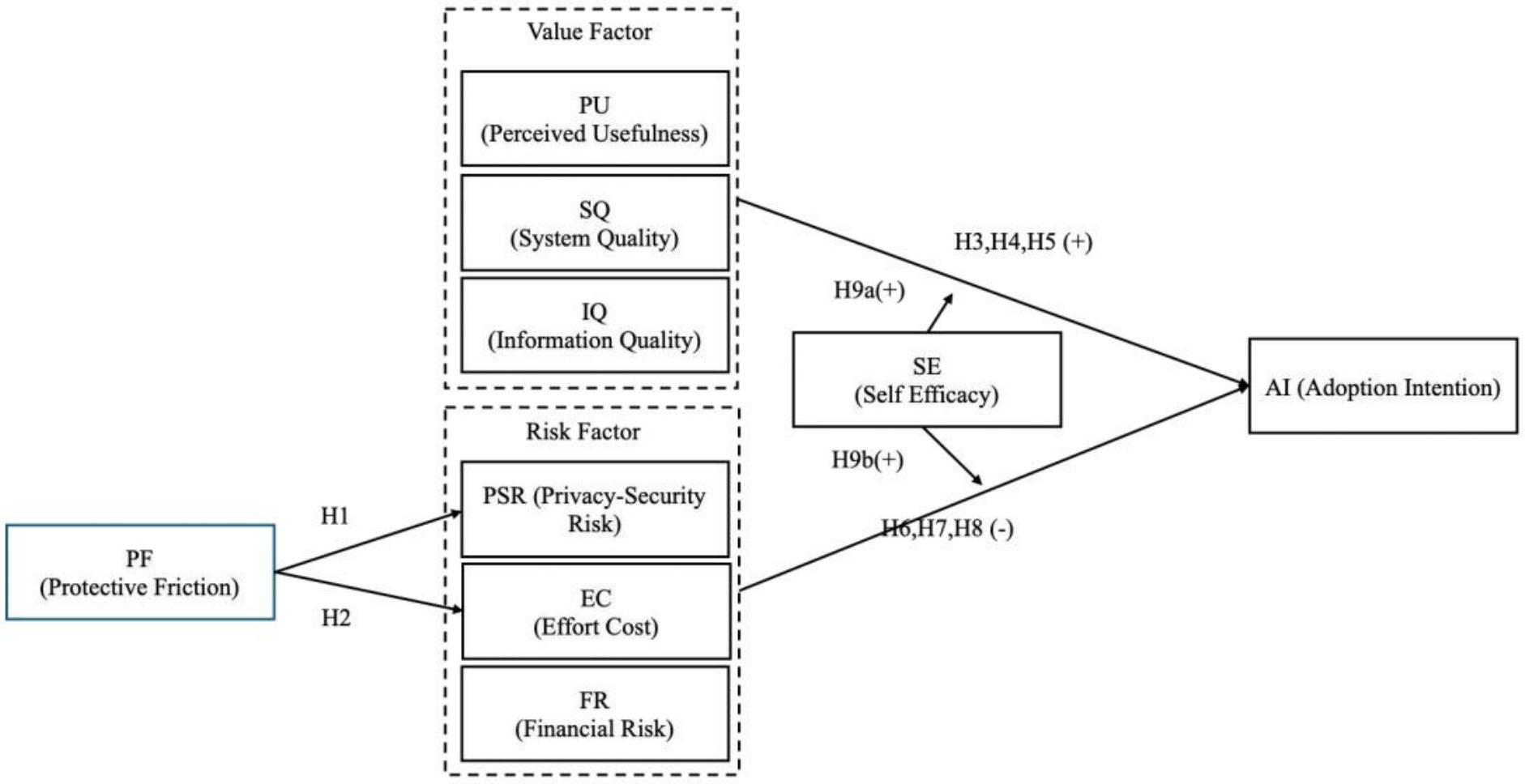

Figure 1: Research Model – Dual-Pathway Model of Protective Friction
Source: This study.

**The Effect of Protective Friction Hypotheses**

The protection pathway predicts that visible security authentication signals the platform's commitment to data protection, thereby alleviating users' concerns about privacy and security breaches. Conversely, the cost pathway predicts that the same authentication steps impose additional time, effort, and cognitive load, raising the perceived cost of using the platform. Protective friction serves a dual purpose: it communicates security commitment while imposing operational costs. The following hypotheses test these opposing pathways.

H1: Protective friction (security authentication) is negatively associated with users' privacy-security risk perception (protection pathway).

H2: Protective friction (security authentication) is positively associated with users' effort cost perception (cost pathway).

**Value Factor Hypotheses**

Consistent with the technology acceptance model and IS success model, we expect that perceived value factors positively influence adoption intention. These factors reflect the usefulness, quality, and accessibility of the platform, dimensions that align with established IS theory.

H3: Perceived usefulness positively affects adoption intention.

Perceived usefulness captures the degree to which users believe the platform delivers valuable credit services. Users who see clear benefits are more likely to adopt despite authentication friction.

H4: System quality positively affects adoption intention.

System quality, including reliability, responsiveness, and technical performance, signals a well-maintained platform, reducing uncertainty and increasing adoption likelihood.

H5: Information quality positively affects adoption intention.

High-quality information (accuracy, completeness, timeliness of credit data) is critical in government credit platforms. Better information quality increases perceived usefulness and adoption. Recent work by Mensah and Mwakapesa (2025) demonstrates information quality's direct impact on e-government service adoption.

**Risk Factor Hypotheses**

Perceived risks are expected to inhibit adoption. However, we propose that the institutional trust buffer, government endorsement and free service provision, neutralizes financial risk. These hypotheses test the risk side of the adoption model, incorporating recent advances in privacy calculus research.

H6: Privacy-security risk perception negatively affects adoption intention.

Users concerned about data breach or unauthorized access are reluctant to adopt credit platform services. This effect is strongest when users perceive weak security, though our protection pathway hypothesis predicts that visible authentication mitigates such concerns. Recent work on privacy calculus in digital government (Huang & Huang, 2023; Meier & Kramer, 2024) confirms this negative relationship.

H7: Effort cost negatively affects adoption intention.

Higher perceived effort, including time required, steps to complete, and cognitive load, reduces adoption intention. Authentication friction directly increases this cost dimension.

H8: Financial risk does not significantly affect adoption intention due to the institutional trust buffer.

Government credit platforms typically operate without direct fees, and users perceive minimal financial risk from free government services. This institutional trust buffer differentiates public-sector adoption from commercial e-commerce contexts.

**Moderation Hypotheses**

The critical test of the comprehension versus ability question lies in the pattern of self-efficacy's moderation across the two pathways. If self-efficacy primarily captures operational ability, it should moderate the effort-cost pathway (H9b supported, H9a not). If it primarily captures comprehension of the friction's protective purpose, it should moderate the risk pathway (H9a supported, H9b not):

H9a: Self-efficacy moderates the relationship between privacy-security risk perception and adoption intention, such that the negative effect of privacy-security risk is weaker for users with higher self-efficacy.

Users with higher technical self-efficacy are better equipped to understand security mechanisms and evaluate risk rationally. They may be less prone to anxiety-driven avoidance when friction signals security commitment.

H9b: Self-efficacy does not significantly moderate the relationship between effort cost and adoption intention.

Effort cost reflects the objective operational burden of authentication steps. Even highly self-efficacious users will experience the same time and cognitive demands; their advantage lies in comprehending the friction's purpose, not in reducing the effort required.

If H9a is supported while H9b is not, this asymmetric pattern would suggest that the mechanism underlying self-efficacy's role in the friction context is comprehension rather than ability, pointing toward a theoretically important but currently unmeasured construct we term protective mechanism comprehension: the degree to which users understand why a security measure exists and how it protects them.

## RESEARCH METHOD

This study employs a convergent mixed-methods design (Creswell & Plano Clark, 2018), combining a quantitative survey with qualitative interviews conducted at a municipal credit center in Shenzhen, China, one of the pilot cities for China's social credit system.

Table 1. Constructs and Measurements

| Construct | Items | Source |
|---|---|---|
| Perceived Usefulness (PU) | 4 items | Davis (1989) |
| System Quality (SQ) | 4 items | DeLone & McLean (2003) |
| Information Quality (IQ) | 4 items | DeLone & McLean (2003) |
| Privacy-Security Risk (PSR) | 3 items | Featherman & Pavlou (2003) |
| Effort Cost (EC) | 3 items | Venkatesh et al. (2003) |
| Financial Risk (FR) | 3 items | Featherman & Pavlou (2003) |
| Self-Efficacy (SE) | 3 items | Compeau & Higgins (1995) |
| Adoption Intention (AI) | 3 items | Venkatesh et al. (2003) |

Source: This study.

### Quantitative Component

A structured questionnaire will be administered to platform users through convenience and snowball sampling. Constructs are measured using established scales adapted to the credit platform context: perceived usefulness (4 items, adapted from Davis, 1989), system quality and information quality (4 items each, adapted from DeLone & McLean, 2003), service accessibility (3 items), privacy-security risk (3 items, adapted from Featherman & Pavlou, 2003), effort cost (3 items, adapted from UTAUT), financial risk (3 items), self-efficacy (3 items, adapted from Compeau & Higgins, 1995), and adoption intention (3 items). All items use 7-point Likert scales. PLS-SEM analysis will be conducted using SmartPLS with 5,000 bootstrap samples, following established guidelines (Hair et al., 2019). The target sample size is 400+ responses to ensure adequate statistical power.

### Qualitative Study

Semi-structured interviews will be conducted with credit center staff members across multiple functional departments (25–30 minutes each). Staff are selected because they combine deep platform knowledge with direct exposure to user feedback, providing a qualitative description of how users experience protective friction. Interviews will be recorded, transcribed, and analyzed using thematic analysis to explore the mechanisms through which users respond to and interpret authentication friction.

## EXPECTED CONTRIBUTIONS AND FUTURE RESEARCH

### Theoretical Contributions

This research is expected to make three theoretical contributions. First, by empirically testing the dual-pathway model, we aim to extend the friction literature, particularly Tomalin's (2023) typology and Distler et al.'s (2020) security-enhancing friction framework, from conceptual classification to empirical mechanism identification. If the dual-pathway model is supported, it would demonstrate that protective friction does not produce a simple positive or negative effect but rather operates through opposing pathways whose net outcome depends on moderating factors.

Second, the asymmetric moderation pattern (H9a vs. H9b) provides a diagnostic test for the comprehension versus ability question. If self-efficacy moderates the risk pathway but not the cost pathway, it would suggest that the critical mediating factor is not users' operational skill but their understanding of why protective friction exists. This would align with Mejtoft et al.'s (2019) finding that satisfaction with friction depends on goal comprehension, and would point toward protective mechanism comprehension as a theoretically important construct. Existing constructs capture adjacent phenomena: self-efficacy measures capability, perceived effectiveness measures belief, and security awareness measures knowledge of threats, but none captures the user's understanding of why a specific protective measure exists and how it functions to safeguard them.

Third, the institutional trust buffer hypothesis (H8) would contribute to understanding how government endorsement alters the traditional risk calculus in technology adoption. If financial risk is indeed neutralized by institutional trust, this has implications for how public-sector digital services should be theorized differently from commercial services.

### Limitations

This research-in-progress faces several limitations that should be addressed in future work. First, the study is limited to a single city (Shenzhen), which constrains generalizability to other regions of China with different institutional contexts or economic development levels. Second, the cross-sectional design prevents causal inference; longitudinal or experimental designs would strengthen claims about the direction of effects. Third, the qualitative component relies on staff interviews rather than direct user interviews, introducing an indirect perspective on user experience that may miss nuances in how users actually perceive and respond to authentication friction. Fourth, protective friction is operationalized through existing constructs (risk perception, effort cost) rather than through direct measurement of the construct itself, potentially missing dimensions unique to protective authentication. Finally, the study's institutional and cultural context, specifically the role of government trust and the unique dynamics of China's social credit system, may limit the applicability of findings to e-government platforms in other countries with different governance structures and user-institution relationships.

**Planned Next Steps**

Beyond the proposed model test, we plan three follow-up activities. First, if the asymmetric moderation pattern is confirmed, we will develop and validate a measurement scale for protective mechanism comprehension through item generation, expert review, and confirmatory factor analysis. Proposed items include: "I understand how facial recognition prevents unauthorized access to my credit data" and "I can explain to someone else why multi-step verification is necessary." Second, we intend to test a mediated moderation model in which self-efficacy's effect on the risk pathway is mediated by protective mechanism comprehension. Third, we plan an experimental study manipulating the transparency of friction's protective purpose (e.g., adding explanatory text during authentication: "This step protects your credit data by..." vs. no explanation) to test whether making the protective logic explicit converts burdensome friction into beneficial friction.

**Practical Implications**

If the proposed model is supported, the findings would suggest a specific design principle for government digital services handling sensitive data: rather than choosing between security and convenience, platform designers should make friction comprehensible. Specifically, this would mean: (1) adding contextual explanations during authentication steps (e.g., "Facial recognition ensures that only you can access your credit report"); (2) implementing risk-tiered verification, lighter authentication for low-risk operations and heavier for high-risk, to reduce unnecessary friction while preserving protection where it matters; and (3) providing differentiated support for users with varying levels of technical understanding.

## ACKNOWLEDGMENT

This work was partially supported by the Provincial Science-Industry-Education Integration Practice Teaching Base (An educational platform for "AI+Management" Disciplines) at Southern University of Science and Technology (SusTech), awarded by the University Teaching Quality and Teaching Reform Program of Guangdong Provincial Education Bureau (ID:SJZLGC202514).The authors would like to acknowledge the partial grant support to the Young Innovative Talent Program of Guangdong (Grant ID: 2023WQNCX067), the DeFin research center, National Center for Applied Mathematics Shenzhen and Shenzhen Key Research Base in Arts & Social Sciences (Intelligent Management & Innovation Research Center of SUSTech).